\documentclass[prl,onecolumn,superscriptaddress]{revtex4}
\usepackage{amsmath}
\usepackage{amsfonts}
\usepackage{amssymb}
\usepackage{graphicx}
\usepackage{bm}
\usepackage{dsfont}

\usepackage[ansinew]{inputenc}
\usepackage{graphicx}
\usepackage{amsmath}
\usepackage{amsthm}
\usepackage{bm}
\usepackage{layout}
\usepackage{float}
\usepackage{txfonts}
\usepackage{amsfonts}
\usepackage{amssymb}%
\begin{document}

\title{In the {\it Kreisgang} between classical and quantum physics}

\author{{\v C}aslav Brukner}

\affiliation{Faculty of Physics, University of Vienna,
Boltzmanngasse 5, A-1090 Vienna, Austria} \affiliation{Institute for
Quantum Optics and Quantum Information, Austrian Academy of
Sciences, Boltzmanngasse 3, A-1090 Vienna, Austria}

\date{\today}

\begin{abstract}

All our statements about the physical world are expressed in terms
of everyday notions and thus in terms of classical physics. This
necessity is behind each of our attempt to extract meaning out of
empirical data and to communicate this knowledge to others. As such,
it must apply also to the account of measurement arrangements and to
the outcome in quantum experiments. On the other hand, however, if
quantum mechanics is universally valid it should be possible to give
a purely quantum mechanical description of objects of increasingly
large sizes, eventually of the measurement devices themselves. It is
suggested to resolve this dilemma by using the method of von
Weizsäcker's circular and consistent movement in a reconstruction
({\it Kreisgang}) in which it is legitimate to recover the elements
with which one started the reconstruction. The parameters entering
the complex amplitudes of a quantum state have an operational
meaning as parameters that specify the configuration of macroscopic
instruments by which the state is prepared and measured. This
classical aspect in each quantum description is at the ``beginning''
of the {\it Kreisgang}. The {\it Kreisgang} is ``closed'' by showing
that under the every-day conditions of coarse-grained measurements a
description of macroscopic instruments emerges in terminology of
classical physics and the three-dimensional ordinary space from
within quantum theory.

\end{abstract}

%\pacs{03.65.-w,03.65.Ta}

\maketitle

\begin{quotation}  {\it Nach Kant ist das wissende Subjekt eben darum nicht als
Substanz zu beschreiben; denn Substanz ist selbst eine Kategorie,
also ein Begriff.}

Carl Friedrich von Weizs\"{a}cker in ``Aufbau der Physik''
\end{quotation}

All our statements about the physical world are statements about the
classical measurement apparatus and its classical features. It does
not make any sense to speak about characteristics of the quantum
system in itself without explicitly specifying the measuring
apparatus. Bohr put much emphasize in his writings on the fact that
the conditions under which an observer can acquire objective
knowledge and communicate it with others requires a classical
description of the measuring apparatus, even then when the phenomena
under investigation are very distinct from those of the classical
world view. For example, in a paper~\footnote{N. Bohr, in {\it
Albert Einstein, Philosopher-Scientist}, edited by P. A. Schlipp
(Library of Living Philosophers, Evanston, IL, 1949), p. 209} of
1949, Bohr stressed that
%\begin{quotation}
``However far the phenomena transcend the scope of classical
physical explanation, the account of all evidence must be expressed
in classical terms. The argument is simply that by the word
'experiment' we refer to a situation where we can tell others what
we have done and what we have learned and that, therefore, the
account of the experimental arrangement and the result of
observation must be expressed in unambiguous language with suitable
application of the terminology of classical physics.''
%\end{quotation}
In similar vein, when discussing the Copenhagen view,
Grunbaum~\footnote{A. Grunbaum, {\it Complementarity in Quantum
Physics and its Philosophical Generalization}, The Journal of
Philosophy, Vol. {\bf 54}, No. 23, American Philosophical
Association Eastern Division: Symposium Papers at the Fifty-Fourth
Annual Meeting, Harvard University, December 27-29, 1957 (Nov. 7,
1957), pp. 713-727.} introduces the ontology of
%\begin{quotation}
macrophysicalism according to which ''the existences of the physical
world are the macro-object and events of classical physics.
Micro-objects do not exit as such at all but can be introduced into
physical theory as a kind of computational or linguistic link
between specifiable experimental arrangements and their observable
consequences. Accordingly, to speak of electrons, for example, as
'interacting' with measuring devices has, existentially, only a
Pickwickian meaning. For only the registrational processes and
components of the classical-describable total experimental
arrangements are existents.''
%\end{quotation}

The decisive role played by the classical measurement apparatus in
the epistemological framework of quantum physics was often not
accepted or even misunderstood by many scholars. This circumstance
was, for example, recently emphasized  by Osnaghi, Freitas and
Freire~\footnote{S. Osnaghi, F. Freitas, and O. Freire Jr., {\it The
origin of the Everettian heresy,} to be published in Studies of
History and Philosophy in Modern Physics (2009).}, who write:
%\begin{quotation}
``Bohr's {\it functional} distinction between object-system and
measuring instrument was replaced by the crude {\it physical}
assumption that macroscopic systems behave classically, which would
introduce an artificial split of the physical world into a quantum
microcosms and a classical macrocosms.''
%\end{quotation}
The main misunderstanding surrounding the alleged dichotomy in the
Copenhagen view comes from the assertion that a measurement device
itself is built up of microscopic systems, such as atoms or protons
and electrons, for which the appropriateness of the quantum
mechanical description seems to be indisputable. In further
elaboration of this view, the measurement procedure is considered to be a
particular kind of physical interaction occurring between system (I)
and measurement apparatus (II) for which it is possible to set up
the quantum state for the apparatus, the Hamiltonian for the interaction
between the system and apparatus, and the Schr{\"o}dinger evolution.
However, from the Copenhagen viewpoint, the quantum state acquires a
physical meaning only when it has been related to the well-specified
measurement procedure through which the observables are defined.
While there is nothing in the theory that would prohibit to reach
the necessary experimental precision to allow a meaningful state
assignment to objects of increasingly large sizes
--- eventually as large as our measurement devices --- it is
indispensable to use classical concepts for describing the
experimental context in which these objects are observed. By
appealing to the logical necessity of making a sharp distinction
between object and the measurement apparatus, the device II can only
be considered as a part of the composite system I+II which is to be
described by the Schr{\"o}dinger evolution, if it loses its previous
status of ``measurement apparatus''. In order to speak about a
quantum state of the composite system I+II, it would actually be necessary to
perform a measurement with another apparatus III by which the
conditions leading to the appearance of quantum features of I+II can
be established, in the same way as a meaningful assignment of
a quantum state to I was acquired through the measurement procedure
defined by apparatus II.

To illustrate the preceding argument take, for example, a spin-1/2
system in the state $|z+\rangle$. What we mean with this syntax or
{\it symbol}, is that the probability to find the outcome ``spin
up'' when the Stern-Gerlach magnet is oriented along an angle
$\theta$ with respect to the $+z$-direction is $p=\cos^2(\theta/2)$.
We see that we never speak about the quantum state {\it per se}, but
always refer to some well-defined configuration of macroscopic
instruments in the ordinary three-dimensional space by which the
conditions defining the probability for an outcome are established.
When considering macroscopic measurement devices -- which in order
to fulfill their function necessarily need to consist of a large
number of elementary constituents -- it may seem natural to assume
that a description of each of its individual constituents and thus
of the entire device can be given purely in quantum mechanical
terms. The main difficulty of such a total reduction of the
description of the measurement apparatus to the quantum description
of its elementary constituents lies in the obvious fact that quantum
states of these constituents can acquire a meaning only in a
classically describable experimental context, thus making the whole
argument indefinite in its hierarchical structure. Criticizing such
attempts Rosenfeld~\footnote{L. Rosenfeld, {\it The Measuring
Process in Quantum Mechanics}, in Supplement of the Progress of
Theoretical Physics, Commemoration Issue for the 30th Anniversary of
the Meson Theory by Dr. H. Yukawa, 223-231, 1965} notes that:
%\begin{quotation}
``... no formalization can be complete, but must leave undefined
some 'primitive' concepts and take for granted without further
analysis certain relations between these concepts, which are adopted
as 'axioms': the concrete meaning of these primitive concepts and
axioms can only be conveyed in a 'metalanguage' foreign to the
formalism of the theory.''
%\end{quotation}
In a similar fashion Peres~\footnote{A. Peres, {\it Quantum Theory:
Concepts and Methods}, Kluwer Academic Publishers, 2002, p.173}
writes:
%\begin{quotation}
``Even if quantum theory is universal, it is not closed. A
distinction must be made between endophysical systems -- those which
are described by the theory -- and exophysical ones, which lie
outside the domain of the theory (for example, the telescopes and
photographic plates used by astronomers for verifying the laws of
celestial mechanics). While quantum theory can in principle describe
anything, a quantum description cannot include everything. In every
physical situation something must remain unanalyzed.''
%\end{quotation}

Though the measurement procedure is not entirely analyzable by means
of quantum mechanical laws, it must be possible, {\it for the
reasons of consistency}, to arrive at an explanation of objective
properties of macroscopic objects as large as apparatuses from
within quantum theory. In a joint work with Johannes Kofler, I have
recently proposed one such approach~\footnote{J. Kofler and {\v C}.
Brukner, {\it Classical World Arising out of Quantum Physics under
the Restriction of Coarse-Grained Measurements}, Phys. Rev. Lett.
{\bf 99} (2007) 180403. J. Kofler and {\v C}. Brukner, {\it
Conditions for Quantum Violation of Macroscopic Realism}, Phys. Rev.
Lett. {\bf 101} (2008) 090403.}. While it is not at variance with
decoherence~\footnote{W.H. Zurek, {\it Decoherence and the
transition from quantum to classical}, Phys. Today {\bf 44}, 36
(1991); W.H. Zurek, {\it Decoherence, einselection, and the quantum
origins of the classical}, Rev. Mod. Phys. {\bf 75}, 715 (2003).},
it differs from it conceptually. It focuses on the limits of
observability of quantum effects of macroscopic objects, i.e., on
the required precision of our measurement apparatuses such that
quantum phenomena can still be observed. We have considered spin
systems and demonstrated that under the restriction of
coarse-grained measurements and the limit of large spins not only
macroscopic realism but even the classical Newtonian laws emerge out
of the Schr\"{o}dinger equation and the projection postulate.
Classical deterministic laws are therefore an effective and
convenient illusion insomuch one's observations are of sufficient
inaccuracy~\footnote{Supported by our every-day experience, this
illusion seems to be very strong and persistent among scholars as
continuous attempts are undertaken in exactly the opposite
direction, namely, towards constructing hidden-variable models which
are supposed to explain probabilistic quantum laws by underlying
deterministic structures. The theorems of Kochen and Specker [S.
Kochen and E. P. Specker, {\it The problem of hidden variables in
quantum mechanics}, J. Math. Mech. {\bf 17}, 59-87 (1967)] and Bell
[J.S. Bell, {\it On the problem of hidden variables in quantum
mechanics}, Rev. Mod. Phys. {\bf 38}, 447 (1966)] show that to
maintain such a view one would need to introduce contextual and
non-local influences.}.

When trying to relate quantum and classical aspects of physical
description in a hierarchical order, we run into the problem of
infinite regress: On one hand, quantum states require classical
instruments to be defined, on the other hand, the instruments
themselves are supposed to be a subject of a quantum-mechanical
description. Von Weisz{\"a}cker considered a complete and
hierarchical description of nature to be unattainable. Instead he
proposed {\it Kreisgang} as a methodological idea of a circular, but
consistent, movement in reconstructing science without a need to
introduce hierarchial relation between its basic
elements~\footnote{C.F. von Weizs{\"a}cker, {\it Aufbau der Physik}
(Carl Hanser, M{\"u}nchen, 1985).}. He
writes~\footnotemark[\value{footnote}]: ``Der Anspruch, damit eine
{\it volle} Beschreibung der Wirklichkeit zu geben, d{\"u}rfte
uneinl{\"o}sbar sein; legitim ist der Anspruch,
eine in der gebenen N{\"a}herung [...] %(der Trennbarkeit der Alternativen)
{\it konsistente} Beschreibung zu geben.'' %In this sense all what we
%can aim for is to give a consistent description of nature within the
%{\it Kreisgang}. it is legitimate
It is important to realize that {\it Kreisgang} is not a circulus
vitiosus of some required deductive ``proof'', rather it is a
consistency argument in which it is legitime and even necessary to
recover the elements of logic with which one started the
reconstruction.

When trying to identify which aspects of the classical description
should be employed at the ``beginning'' and recovered at the ``end''
of the {\it Kreisgang} in the reconstruction of quantum theory, it
is instructive to analyze Bohr's writing~\footnote{N. Bohr, {\it
Quantum Physics and Philosophy. Causality and Complementarity}. In
Philosophy in Mid- Century: A Survey. R. Klibansky, ed. Florence: La
Nuova Italia Editrice, 1958.} from 1958: ``In actual experimental
arrangements, the fulfillment of such requirements [describing
unambiguously the apparatus and results of measurement] is secured
by the use, as measuring instruments, of rigid bodies sufficiently
heavy to allow a completely classical account of their relative
positions and velocities. In this connection, it is also essential
to remember that all unambiguous information concerning atomic
objects is derived from the permanent marks -- such as a spot on a
photographic plate, caused by the impact of an electron -- left on
the bodies which define the experimental conditions. Far from
involving any special intricacy, the irreversible amplification
effects on which the recording of the presence of atomic objects
rests rather remind us of the essential irreversibility inherent in
the very concept of observation. The description of atomic phenomena
has in these respects a perfectly objective character, in the sense
that no explicit reference is made to any individual observer and
that therefore [...] no ambiguity is involved in the communication
of information.'' As noted by Howard~\footnote{D. Howard, {\it What
Makes a Classical Concept Classical? Toward a Reconstruction of
Niels Bohr's Philosophy of Physics}, electronic version at
http://www.nd.edu/ndhoward1/classcom.pdf}, Bohr distinguishes here
two classical aspects of measurement apparatus: (i) the occurrence
of ``irreversible amplification effects'' such as a spot on a
photographic plate, a ``click'' in a photo-detector or the pointer
moving in a given position, and (ii) ``the use, as measuring
instruments, of rigid bodies sufficiently heavy to allow a
completely classical account of their relative positions and
velocities.'' %For example, the orientation of the Stern-Gerlach
%magnet as well as the location of the spot at the observation screen
%are crucial for the determination which of the two outcomes ``spin
%up'' or ``spin down'' are realized in the measurement.
Three comments are appropriate at this point. Firstly, as pointed
out by B\"{a}chtold~\footnote{M. B\"{a}chtol, {\it Are all
measurement outcomes ``classical''?} Studies in History and
Philosophy of Modern Physics {\bf 39} (2008) 620– 633.},  Bohr never
argued that all measurement outcomes must allow a full classical
explanation in which the outcomes are understood as revealing some
properties of the world existing prior to and independent of
measurements (i.e. in terms of what we today call hidden variables).
On the contrary, he repeatedly stressed that in quantum mechanics it
is impossible to describe the measurement outcomes by applying
simultaneously mutually complementary notions. Secondly, the
investigations in the foundations of quantum mechanics have shown
that not the ``size'' or ``mass'' of an object is relevant for the
transition to classicality~\footnote{A. Garg and N.D. Mermin, {\it
Bell inequalities with a range of violation that does not diminish
as the spin becomes arbitrarily large}, Phys. Rev. Lett. {\bf 49},
901 (1982)},[5,6], but rather the precision of our observations.
Therefore, Bohr’s arguments concerning the necessity of using, ``as
measuring instruments, rigid bodies sufficiently heavy to allow a
completely classical account of their relative positions and
velocities'' is correct insofar as it is increasingly demanding to
perform accurate measurements on such bodies. Thirdly, one should
leave open the possibility that explicit examples of what we today
perceive as ``irreversible amplification effects'' might tomorrow be
manipulated in a quantum coherent and reversible way. Clearly, to
make an experimental record confirming this coherence one would
again need to have some ``irreversible amplification effects''.
While, therefore, establishment of measurement documents is a
necessary condition to extract meaning out of the documents
whatsoever, the cut between the measurement apparatus and the object
under observation can be shifted and is, perhaps, only conditional
on the current status of technological development.

The purpose of this manuscript is to show -- on the example of
quantum theory of spin -- that the classical aspects of observation
can be employed in the reconstruction of quantum theory in a
consistent way, thus fulfilling the condition for a {\it Kreisgang}.
To this end, I propose the following {\it consistency condition}: If
in a theory the ``state'' of a (directional) elementary system
(spin) requires $d$ real parameters to be specified completely, then
macroscopic objects and instruments under the restriction of
coarse-grained observations must allow for an ``objective''
description of both (a) the account of the experimental arrangement
and (b) the experimental outcome in a $d+1$-dimensional (ordinary)
space. This is for the simple reason that {\it operationally} real
parameters specifying the state {\it are} the parameters determining
the configuration of macroscopic instruments by which measurements
on the state are performed. Note, however, that {\it a priori}, the
consistency argument does not determine the dimension $d$. For
complex quantum mechanics $d=2$, for quaternionic quantum
mechanics~\footnote{P.B. Slater, {\it Bayesian inference for complex
and quaternionic two-level quantum systems}, Physica A, {\bf 223}, 1
(1996), pp. 167-174(8).} $d=5$. It is legitimate to think that
starting with the theory that differs from complex quantum theory
and going into the limit of coarse-grained measurement one might
arrive at the ``classical physics'' embedded in a space of
dimensions different than the one of our everyday life. It should
also be clarified that the ``objective'' description in the
consistency condition given above is to be understood as a
description that is communicable in an unambiguous manner, that is,
independent of any observer, {\it not} as a description that allows
explanation in terms of hidden variables.

I now show how the {\it Kreisgang} can be achieved in the
reconstruction of (complex) quantum theory of spin. Consider a
quantum-mechanical system of a large dimension with a set of $2j+1$
distinguishable (orthogonal) states $|\triangledown\rangle,
|\spadesuit\rangle, ..., |\star\rangle$. The chosen notation should
indicate that {\it a priori} no ordering of the states has been
assumed. I suppose that the outcomes of everyday measurements are in
principle directly observable with our sensory organs (e.g. with our
eyes) such that the resolution of the measurements is not sharp,
making it impossible to resolve individual states, but bunching
together a number of outcomes into ``slots''. It is thus the context
of a measurement where the notion of ``neighboring'' states emerges,
by treating those states as close which correspond to outcomes that
are observed close in real space. I therefore choose an arbitrary
set of $2j+1$ orthogonal states from the Hilbert space and impose on
them an ordering: $|\triangledown\rangle \equiv |m=-j\rangle,
|\spadesuit\rangle \equiv |m=-j+1\rangle, ..., |\star\rangle \equiv
|m=j\rangle$, which is to be understood as identifying them with the
ordered outcomes in real space. The full justification of this
procedure can only be achieved in the {\it Kreisgang} when showing
the consistency of the choice of the state ordering.

Consider now a measurement of the eigenvalues $m$ of the observable
defined as $\hat{J}_{z} = \sum_{m=-j}^{j} m \vert m \rangle \langle
m \vert $ in units where $\hbar=1$. At this stage of the argument one could think about this
as purely mathematical definition of an observable, but bear in mind
that, physically, it represents the $z$-component of the spin
observable. I introduce the spin coherent states $|\Omega \rangle
\equiv |\vartheta,\varphi \rangle =\sum_{m=-j}^{m=j}
\binom{2j}{j+m}^{1/2} \cos^{j+m}(\vartheta/2)
\sin^{j-m}(\vartheta/2) e^{im \varphi}|m\rangle,$ with polar
$\vartheta$ and azimuthal angle $\varphi$, which are the eigenstates
with maximal eigenvalue of a spin operator $\hat{\mathbf{J}}_{\Omega
} $ pointing into the direction $\Omega \equiv(\vartheta,\varphi)$:
$\hat{\mathbf{J}}_{\Omega }\left\vert \Omega\right\rangle
=j\left\vert \Omega\right\rangle $. It is
known that any spin-$j$ state can be written in the quasi-diagonal form $\hat{\rho}=%
%TCIMACRO{\tiint }%
%BeginExpansion
{\textstyle\iint}
%EndExpansion
P(\Omega)\,|\Omega\rangle\langle\Omega|\,$d$^{2}\Omega$ with
d$^{2}\Omega$ the solid angle element and $P(\Omega)$ a normalized
and \textit{not necessarily positive} real function~\footnote{F.T.
Arecchi, E. Courtens, R. Gilmore, and H. Thomas, {\it Atomic
Coherent States in Quantum Optics}, Phys. Rev. A {\bf 6}, 2211-2237
(1972).}.

As noted above, the resolution of our everyday measurements is not
sharp, but bunches together $\Delta m$ outcomes of the spin
$z$-component $\hat{J}_{z}$ into "slots" $\bar{m}$. There are
$(2j+1)/\Delta m$ different slots. It was shown [6] that if the
measurement coarseness is much larger than the intrinsic uncertainty
of coherent states, i.e. $\Delta m\!\gg\!\!\sqrt{j}$, then the
probability for getting the particular slot $\bar{m}$ can be
computed via integration of a {\it positive} probability
distribution $Q$ as given by
\begin{equation}
\mbox{prob}(\bar{m})=\int_{\Omega_{\bar{m}}} Q(\Omega) dQ.\label{Q}
\end{equation}
Here $Q = \frac{2j+1}{4\pi} \langle \Omega\vert \hat{\rho} \vert
\Omega \rangle$ is the well-known $Q$-function in quantum
optics~\footnote{G.S. Agarwal, {\it Relation between atomic
coherent-state representation, state multipoles, and generalized
phase-space distributions}, Phys. Rev. A {\bf 24}, 2889 (1981); G.S.
Agarwal, {\it Perspective of Einstein-Podolsky-Rosen spin
correlations in the phase-space formulation for arbitrary values of
the spin}, Phys. Rev. A {\bf 47}, 4608 (1993).} and $\Omega_m$ is
the angular section of polar angular size $\Delta\Theta_{\bar{m}}
\sim \Delta m \gg 1/\sqrt{j}$ whose projection onto the $z$ axis
corresponds to the slot $\bar{m}$. The quantum description of a spin
in a general (mixed) state requires an order of $j^2$ real numbers
to be specified. However, the effective description under
coarse-grained measurements is always that of a convex mixture over
classical spins embedded in the ordinary three-dimensional space
(i.e. whose orientation requires two polar angles to be defined)
independently of the spin size. In this way the {\it Kreisgang} is
closed. We began with a description of a spin-1/2 quantum state in
terms of two polar angles and we end with a description of large
systems under coarse-grained measurements in terms of the two polar
angles. This is not surprising if one accepts that parameters in the
state $|\psi \rangle \equiv \cos(\vartheta/2)e^{-i\varphi/2}
|z+\rangle +\sin(\vartheta/2)e^{+i\varphi/2} |z-\rangle$ of a
spin-1/2 system have no other meaning than that of the angles along
which the macroscopic Stern-Gerlach magnet needs to be oriented to
observe the definite outcome ``spin-up''. We conclude that the
account of the experimental arrangement is given in terms of the
three-dimensional space (condition (a)).

Upon a coarse-grained measurement with outcome $\bar{m}$, the state
$\hat{\rho}$ is reduced to $\hat{\rho}_m$ and will change (this can
be seen in the change of the $P$-function) if $\hat{\rho}$ contains
coherence terms (i.e. non-diagonal elements) across the slots.
Nonetheless, it can be shown [6] that the $Q$ distribution before
the measurement is the weighted mixture of the distributions
$Q_{\bar{m}} =\frac{2j+1}{4\pi} \langle \Omega |
\hat{\rho}_{\bar{m}} | \Omega \rangle$ of the possible reduced
states, i.e.
\begin{equation}
Q(\Omega)= \sum_{\bar{m}} \mbox{prob}(\bar{m}) Q(\Omega_{\bar{m}}).
\end{equation}
This shows that a coarse-grained measurement can be understood
classically as reducing the previous ignorance about predetermined
properties of the spin system. For example, the Schr\"{o}dinger-cat
state $|\Psi\rangle = (1/{\sqrt{2}}) (|j+\rangle + |j-\rangle)$ will
change into a classical mixture $\hat{\rho}_{mix} = (1/2) (|j+
\rangle \langle j+| + |j- \rangle \langle j-|)$ under a
(non-selective) coarse-grained measurement (e.g. of $\hat{J}_z$),
but the two states will be perceived as the very same macroscopic
state because the respective $Q$ functions differ only in
exponentially small (in $j$) amount. Therefore, the set of all
coarse-grained measurements is non-invasive~\footnote{For the
rigorous definition of non-invasiveness see: A.J. Leggett and A.
Garg, {\it Quantum mechanics versus macroscopic realism: Is the flux
there when nobody looks?}, Phys. Rev. Lett. {\bf 54}, 857 (1985).
A.J. Leggett, {\it Testing the limits of quantum mechanics:
motivation, state of play, prospects}, J. Phys. Condens. Matter {\bf
14}, R415 (2002).} on that part of state description that is
relevant for predictions of the coarse-grained measurements, namely
for the $Q$ function.

For example, if we remove a single atom from a Stern-Gerlach magnet
or flip the state of one of its spins, the quantum state of the
entire magnet will change into an orthogonal one, but at the
macroscopic level we perceive it as the very same Stern-Gerlach
magnet. This clearly shows that quantum and classical notions of
distinguishability are different. The information represented by the
$Q$-function can be read out repeatedly and independently by
different observers, and it can be copied and communicated, without
disturbing it. As such, it represents the ``classical'' part of the
quantum description. The fact that the experimental outcome, as
represented by a macroscopic property of measuring instrument, can
directly be observed by means of the observer's sensory organs and
will not change under this observation remaining accessible to other
observers, enables inter-subjectivity, i.e. it enables the observers
to agree on the meaning of the outcome. This explains why the
experimental outcome can be given certain level of ``objectivity''
(condition (b)).

As stressed before, the fact that the measurement outcome is
experienced as a stable~\footnote{If the spin-$j$ is realized as the
a set of $N$ spins of length 1/2, then a change of the entire state
in the number of spins in the order of $\sqrt{N}$ will not change
the macroscopic property of the state.} macroscopic property which
can be repeatedly read out by independent observers, thus enquiring
 a level of ``objectivity'', does not imply that these macroscopic
properties of the measurement apparatus can be understood as
existing prior to and independent of the choice of the entire
measurement context. For example, in the case of the measurements of
different observables with two possible outcomes $a_1$ and $a_2$, no
dynamical equations of classical physics or even of any local
realistic theory is able to determine into which of the two
positions $x_1$ or $x_2$ the pointer of the apparatus will evolve
from its initial position in all possible measurements, where $x_1$
is the position corresponding to outcome $a_1$ and $x_2$ is the
position corresponding to $a_2$. This is signified by the fact that
two separated parties, performing only coarse-grained local
measurements on the subsystems of an entangled quantum system can
nonetheless violate Bell’s inequality~\footnote{W. Son, J. Kofler,
M.S. Kim, V. Vedral, and {\v C}. Brukner, {\it Positive phase space
transformation incompatible with classical physics}, Phys. Rev.
Lett. \textbf{102}, 110404 (2009).}, demonstrating the conflict with
local realism. The important point here is that although for {\it
fixed} measurement settings on the two sides of the Bell experiment,
the results of local coarse-grained measurements can have a
classical description in terms of the $Q$-functions, similarly as
indicated by Eq.~(\ref{Q}), i.e. can have a local hidden-variable
model, the local changes of the settings in the two local
laboratories {\it cannot be explained} by local transformation of
the local variables in the hidden-variable model.

In conclusion, while it is unavoidable that certain aspects of
classical description enter the reconstruction of quantum theory,
this must be performed in a consistent manner -- following the
requirement of von Weizsäckers circular movement of knowledge known
as {\it Kreisgang}.  I began the {\it Kreisgang} with a
quantum-mechanical description of an elementary (spin) system. The
parameters entering the complex amplitudes of the quantum state
operationally have the meaning of the parameters that specify the
configuration of macroscopic instruments in the three-dimensional
(ordinary) space (for example, the orientation of the Stern-Gerlach
magnet). I closed the {\it Kreisgang} by showing that under
every-day coarse-grained observations, one arrives at a description
of macroscopic instruments in the terminology of classical physics
and three-dimensional ordinary space from within the quantum
description of its elementary quantum constituents.

I am grateful to M. Aspelmeyer,  J. Kofler, and S. Osnaghi for
discussions. I acknowledge financial support from the Austrian
Science Fund (FWF) and the Foundational Questions Institute (FQXi).
%%%%%%%%%%%%%%%%%%%%%%%%%%%%%%%%%%%%%%%%%%%%%%%%%%%%%%%%%
%%%%%%%%%%%%%%%%%%%%%%%%%%%%%%%%%%%%%%%%%%%%%%%%%%%%%%%%%

\end{document}